\documentstyle[12pt,a4,epsfig]{article}
\newcommand{\beq}{\begin{equation}}
\newcommand{\eeq}{\end{equation}}
\newcommand{\beqar}{\begin{eqnarray}}
\newcommand{\eeqar}{\end{eqnarray}}
\newcommand{\p}{^{\prime}}

\newcommand{\D}{\partial}

\newcommand{\ep}{\varepsilon}

\newcommand{\PR}{Phys. Rev.\ }

\newcommand{\PL}{Phys. Lett.\ }
 \newcommand{\NP}{Nucl. Phys.\ }

\begin{document}
\vspace*{-2cm}

\begin{flushright}
NTZ 09/2002
\end{flushright}

\begin{center}
{\LARGE \bf  Parton interactions in the  Bjorken asymptotics
}\footnote{\it Contribution to the 36th Winter School of
St. Petersburg Nucl. Phys. Institute} \\[2mm]

\vspace{1cm}
R. Kirschner
 \vspace{1cm}

 Institut f\"ur Theoretische Physik, Universit\"at Leipzig,
\\
Augustusplatz 10, D-04109 Leipzig, Germany

\end{center}

\vspace{1cm}
\noindent{\bf Abstract:}
We demonstrate the effective action scheme for the leading parton
interactions and discuss the symmetry properties.
The interaction kernels are
particular cases of conformal symmetric two-particle kernels. There is a
direct relation to the conformal symmetric rational solution of the
Yang-Baxter equation.

%\vspace*{\fill}
%\eject
\vspace{1cm}

\section{Introduction}
\setcounter{equation}{0}

During three decades high energy  processes related to the
Bjorken limit of scattering amplitudes are playing a major role in the
investigation of the hadronic structure and the interaction of hadronic
constituents. Quite a number of theoretical concepts and methods have
been developed resulting by now in a standard treatment described in
textbooks \cite{JCPS} and reviews \cite{GStBLKod}.

In the last years increasing attention is devoted to topics going beyond
the standard deep inelastic structure functions, among them the long
standing problem of the explicite treatment of higher twist contributions
and their scale dependence \cite{twist}, 
the generalization to non-forward kinematics
\cite{SPD}, interpolating between the scale dependence of structure
functions (DGLAP evolution \cite{DGLAP})  and the  scale dependence of
light cone meson wave functions (ERBL evolution  \cite{ERBL} ) as two 
 limiting cases. The existence of this interpolation 
has been pointed out early in \cite{GR}.

The small $x$ behaviour of structure functions and related questions
have drawn the attention also to the relations between the Bjorken and the
Regge limits. The evolution in the latter asymptotics, as far as it
proceeds in the perturbative region, is represented by the BFKL equation
\cite{BFKL,Lev89,Levrep}.

In view of these topics a discussion of the Bjorken asymptotics invoking
concepts not aligned to the standard approach may be of interest.
We advocate an effective action approach \cite{DKBj} in analogy to the one
developed for the Regge asymptotics \cite{effaction}.

The light ray position as the essential variable of operators in the
Bjorken limit appeared explicitely first in the approaches by Geyer,
Robaschik et al. \cite{GR} and by Balitsky and Braun \cite{BB}.

The parton interaction operators can be obtained  as
external field effective vertices, where two type of external fields
are introduces describing the asymptotic interaction with the
currents of high virtuality and with the hadrons. This calculation can be
done analyzing the space-time field configurations only without 
transforming to momentum space.

From the very beginning conformal symmetry was underlying the ideas
about Bjorken scaling and its violation. It turned into a tool for
finding multiplicatively renormalized operators \cite{ERBL,Makeenko} and
for
relating the forward to the non-forward evolutions \cite{BuFKL,Muller}.
Combining conformal and supersymmetry leads to interesting relations
between different evolution kernels \cite{BuFKL}.

Integrability of the effective interaction in high energy QCD amplitudes
has been discovered by Lipatov \cite{LevPadua} in the Regge limit for the
case of multiple exchange of reggeized gluons and a similar
symmetry property  was expected in the Bjorken limit. The first examples
of higher twist evolutions tractable by integrability have been
studied by Braun, Korchemsky et al. \cite{BDKM} and by Belitsky
\cite{AB}. 
Relying on well known methods \cite{QIS} solutions of the Yang-Baxter
equation with conformal ($sl(2)$ and superconformal ($sl(2|1)$)
symmetries have been ontained \cite{DKK} in a form suitable 
for the Bjorken asmptotics. Some results will be reviewed in sections 3 and
4.

 \section{Parton interactions in QCD}
\setcounter{equation}{0}

The standard process is electron - proton scattering $e p \rightarrow e X $
at high energy $\sqrt s $ with a large momentum transfer $q, -q^2 = x s, x =
{\cal O} (1) $, from the electron to the hadron. The hadronic part of the
process can be written in terms of the forward imaginary part of the virtual
photon - proton amplitude.

 The non-forward deeply virtual Compton
scattering off the proton , $\gamma^*(q_1) + p(p_1) \rightarrow
\gamma^*(q_2) + p(p_2) $, is an appropriate generalization (Fig. 1a):
\beqar
q_{1/2} = q^{\prime} - x_{1/2} p^{\prime}, \ \
q^{\prime 2} = 0, p^{\prime 2} = 0, p^{\prime } \approx p_1, s = 2
p^{\prime }  q^{\prime } , \\ \nonumber
-q_{1/2}^2 = Q_{1/2}^2 = x_{1/2} \ \ s.
\eeqar
The Bjorken limit corresponds to $s \rightarrow \infty $ with the
Bjorken variables $x_1$ and/or $x_2$ being of order $1$.

%\hspace*{1cm}

\begin{figure}[htb]
\begin{center}
\epsfig{file=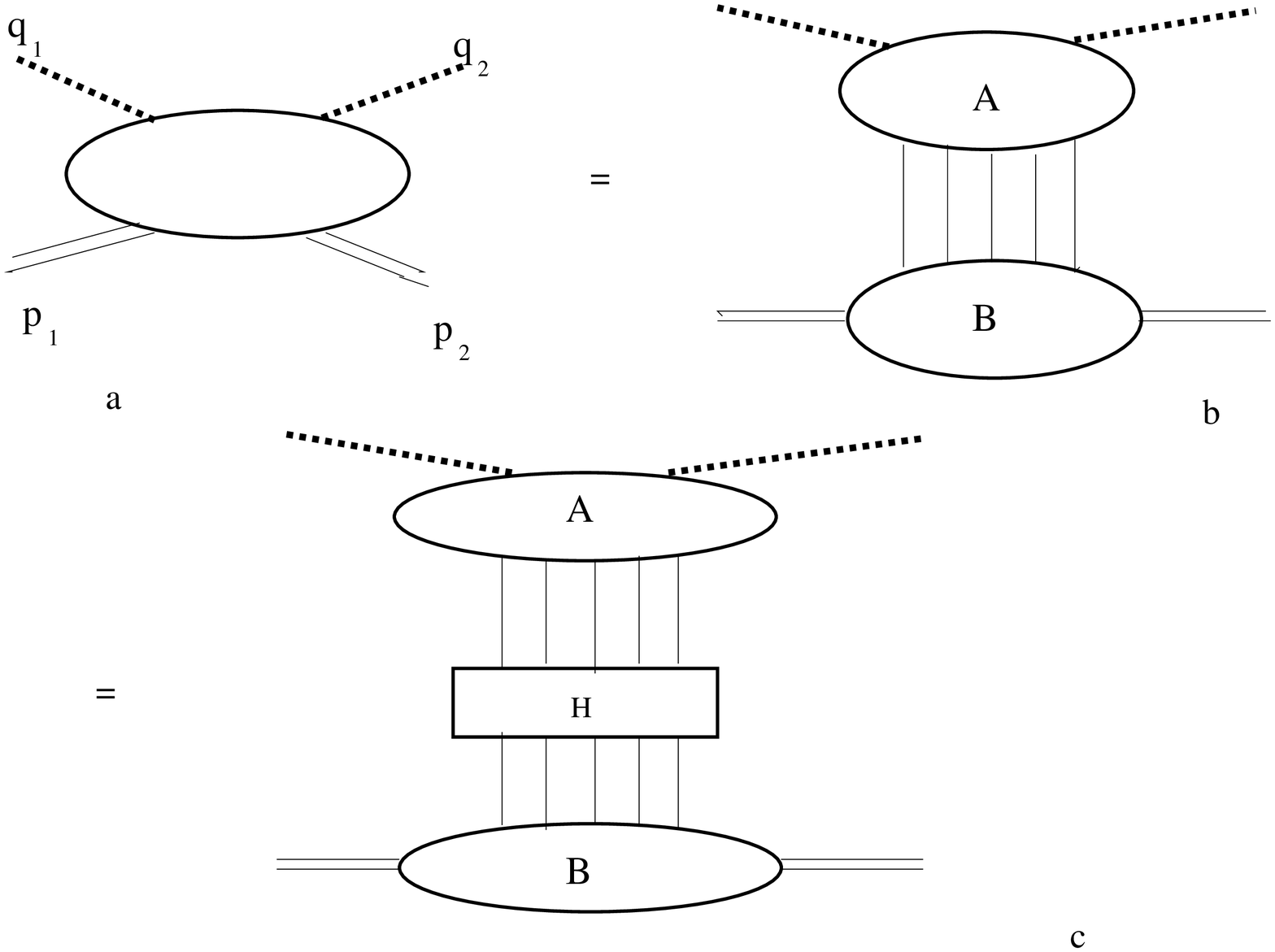,width=10cm}
\end{center}
\vspace*{0.5cm}
\caption{\small
Large-scale factorization of the amplitude. A summation over t-channel
intermediate states is implied.
}
\end{figure}

Let us represent the light cone factorization graphically as in Fig. 1b where
a sum over t-channel parton intermediate states is implied.
The approach to the asymptotics can be visualized as a process of
iterating this factorization. The next step is shown in Fig. 1c.

 The
Green function H (not necessarily connected) represents the parton
interaction. The effective action is to describe just this parton
interaction.

In the lowest order of perturbative expansion there are only pair
interactions. The partons can be identified as modes of the underlying
gluon $A$ and quark $ f, \tilde f$ fields in the gauge $q^{\prime \mu}
A_{\mu} = 0$ after integrating over the redundant field components
$p^{\prime \mu} A_{\mu}$ and $p^{\prime \mu} \gamma_{\mu} \ \psi $.

We represent 4-vectors $x^{\mu} $ by their light cone components
$x_{\pm} $ and a complex number involving the transverse
components $ x_{\perp} = x_1 + i x_2$.
 In the case of the gradient vector
we change the notation in such a way to have $\partial_+ x_- =
\partial_- x_+ = 1, \partial_{\perp} x = \partial_{\perp}^* x^* = 1$.
In particular the complex valued field $A$ represents the transverse
components of the gauge field. We choose the frame where the light-like
vector $q^{\prime}$ has the only non-vanishish component
$q^{\prime}_- = \sqrt {s/2}$ and $p^{\prime }$ the only non-vanishing component
$p_+^{\prime } = \sqrt {s/2}$.

We decompose the Dirac fields into light cone components,
\beqar
\psi = \psi_- + \psi_+, \ \ \gamma_- \psi_+ = \gamma_+ \psi_- = 0 \cr
\psi_+ = f u_{+-} + \tilde f u_{++}, \ \ \ \gamma^{\mu } = {2 \over s}
(\gamma_- q^{\prime \mu } + \gamma_+ p^{\prime \mu } ) + \gamma^{\mu }_{\perp }.
\eeqar

$u_{a,b}, \ \ a,b = \pm, $ is a basis of Majorana spinors,
\beq
\gamma_+ u_{-,b} = \gamma_- u_{+,b} = 0,  \ \
\gamma u_{a, -} = \gamma^* u_{a, +} = 0.
\eeq
The gauge group structure of the action has been written by using
brackets combining two fields into the colour states of the adjoint ($a$)
and  of the two fundametal ($\alpha$ and $*\alpha$) representations:
\beqar (A_1^* T^a A_2 ) = -i f^{a b c } A_1^{* b} \ A_2^c, \
     (f_1^* t^a f_2) = t^a_{\alpha \beta} f^{* \alpha}_1 f_2^{\beta}, \cr
      (f^* t^{\alpha} A ) = t^c_{\beta \alpha} f^{* \beta} A^c, \
     (A^* t^{* \alpha } f) = t^b_{\alpha \beta} A^{* b} f^{\beta }.
\label{brackets}
\eeqar
We shall restrict the detailed discussion to pure gluodynamics,
presenting the final results with fermions included.

The action in light-cone form can be recovered from the
kinetic term
\beqar
\label{kinetic}
\int d^4x -2 A^{*a} (\D_+ \D_- - \D^{\perp} \D^{\perp *} ) A^{a} \cr
\ = \  \int d^4x -2 A^{*a} \D_+ \D_-A^{a}   -
\D^{\perp} \D A^{* a} \D^{-2}  \D^{\perp *} \D A^{a} 
\eeqar
by extending the transverse derivatives in the second form of 
eq. (\ref{kinetic}),   relying on the residual gauge
symmetry,
\beq
\label{extend}
\D^{\perp} A^{a} \rightarrow ({\cal D}^{\perp } A)^{a} =
\D^{\perp } A^{a} + \frac{ig}{2} (A^* T^{a} A),
\eeq
The result can be written as
\beqar
S \ = \ & \int d^4 x  \{
 -2 A^{a *} (x) (\partial _+ \partial_- - \partial_{\perp} \partial_{\perp}^* )
A^{a} (x) \ \ \ \ \ \ \ \ \  \cr
 & + \frac{g}{2} ( \partial_1 \hat V^*_{123} [ \partial_1 A^a (x_1) (A^*(x_2) T^a
A^*(x_3)) + {\rm c.c.})  \cr
 + \frac{g^2}{4} & \hat V_{1 1^{\prime}, 2 2^{\prime} }
[ 2 (A^*(x_1) T^c \partial A(x_{1^{\prime}})
(\partial A^*(x_2) T^c A(x_{2^{\prime}}) ) 
 ]_{x_i = x_i^{\prime } = x } \     \}.
\label{action}
\eeqar
The elimination of redundant field components has lead to non-local
vertices,
\beqar
\hat V^*_{123} = {i \over 3 \partial_1 \partial_2 \partial_3} [
\partial_{\perp 1}^* (\partial_2 - \partial_3) + {\rm cycl.} ], \ \
\hat V_{ 1 1^{\prime }, 2 2^{\prime }} = (\partial_1 +
\partial_{1^{\prime }})^{-2}.
\eeqar
Here and in the following we omit the space index $+$ on derivatives, i.e.
derivative operators not carrying subscripts $-, \perp $ are to be read as
$\partial_+$.  Integer number subscripts refer to the space point
on which the derivative  acts. The definition of the inverse $\partial^{-1}$ is to be
specified.

\section{Space-time picture}
\setcounter{equation}{0}

The  block H of
the virtual Compton amplitude  describes the interaction between
sources located in the vicinity of the light ray $ x_{\perp} = 0, x_- =
0, x_+ = z \in {\bf R}^1 $ and other sources the distribution of which
is almost constant in the variables $x_{\perp}^{\prime}, x_+^{\prime}$
depending essentially only on the coordinate along the light-ray
$x_-^{\prime } = z^{\prime} $. The small width of the first distribution
near
the light cone is characterized by the short  distance scale $\Delta
\sim Q^{-1} $ and
the  variation of the second distribution in directions away from the
light ray is determined by the large distance scale $\delta \sim
m^{-1}$.

Consider now the QCD functional integral with such sources or the
related  vertex functional with corresponding external gluon and quark
fields.
 we divide
 the fields of quarks and gluons into two types of external fields
$A^{(\pm) }$ and a quantum fluctuation $A_q$,
\beq A \rightarrow A^{(+)} + A_q + A^{(-)}.
\eeq
$A^{(+)} $ has the support near the light cone and has to be substituted
by the following expression in terms of (regularized) delta functions,
\beq A^{(+)} (x) = A_1 (z) \delta (x_+) \delta^{(2)} (x_{\perp}) +
{\cal O} (\Delta  ).
\label{fieldasymp+}
\eeq
The other external field $A^{(-)}$ has to be substituted as
\beq A^{(-)} (x^{\prime}) \  = \ A_1^{\prime } (z^{\prime}) \ {\rm const}
+
 {\cal O}( m ).
\label{fieldasymp-}
\eeq
The vertex functional or the external field effective action is now
obtained by doing   the integration over the quantum fluctuations $A_q$.
Consider in particular the resulting vertex involving two $A^{(+)}$ and
two $ A^{(-)}$ type fields on the tree level. It has the form
\beqar
\int d^4x_1 d^4 x_2  d^4x_{1^{\prime }} d^4 x_{^{\prime }}
A^{(+)}(x_1)  A^{(+)} (x_2)      G(x_{1 1^{\prime }} )
G(x_{2 2^{\prime}} )\cr
[\tilde V_{1 1^{\prime }} G(x_{1^{\prime} 2^{\prime }}) \tilde V_{2
2^{\prime }} + \delta^{(4) } (x_{1^{\prime} 2^{\prime }} )
V_{ 1 1^{\prime }, 2 2^{\prime }} ]  A^{(-)} (x_{1^{\prime} }) A^{(-)}
(x_{2^{\prime }})
\label{vertexansatz}
\eeqar

Contributions from disconnected graphs have to be added; they result in the
cancellation of IR divergencies.

$\tilde V_{1 1^{\prime }}, \tilde V_{2 2^{\prime }} $ are simply related
to the triple vertex in (\ref{action}) depending on the
case considered and  $G(x)$
stands for the quark or gluon propagator.

\begin{figure}[htb]
\begin{center}
\epsfig{file=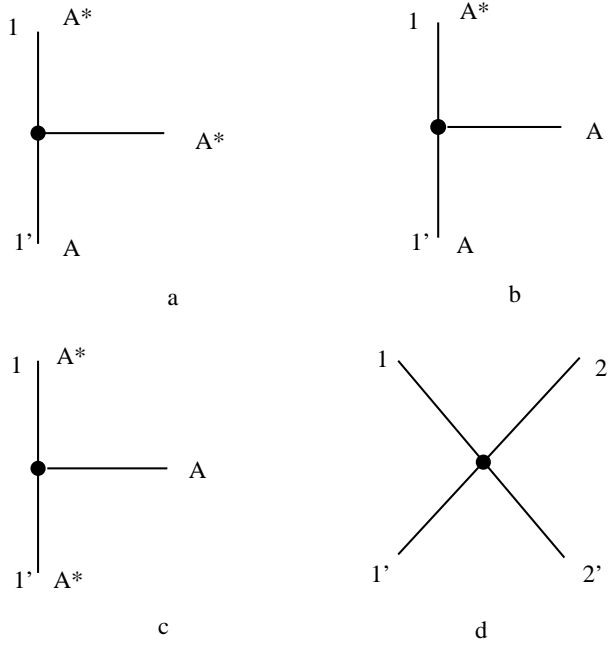,width=8cm}
\end{center}
\vspace*{0.5cm}
\caption{\small
Triple and quartic vertex graphs.
}
\end{figure}
For example, the triple vertex contributes to the tree graph
Fig. 2a as
\beq
-ig \int d^4x_1 x^4x_{1\p} \frac{\D^{2 \prime}}{\D_1 + \D_{1\p} }
\left (\frac{\D_1^{* \perp}}{\D_1} - \frac{\D_{1\p}^{*\perp }}{\D_{1\p}}
\right ) A_q^* (A_1^* T^{a} A_{1\p} ) |_{x_q = x_{1\p}}
\eeq
and reduces in application to $A^{(+)} (x_1), A^{(-)} (x_{1\p}) $
to
\beq
\tilde V_{1 1\p} =
-ig \int d^4x_1 x^4x_{1\p} \frac{\D^2_{1\p}}{\D_1 + \D_{1\p} }
\frac{\D_1^{* \perp}}{\D_1}
 A_q^* (A_1^* T^{a} A_{1\p} ) |_{x_q = x_{1\p}}
\eeq

 We
substitute the particular asymptotic form of the external fields
(\ref{fieldasymp+}, \ref{fieldasymp-}) and obtain
\beqar
c \ln \frac{Q^2}{m^2} \  
\int dz_1 dz_2 dz_{1^{\prime }} dz_{2^{\prime }}
A_1 A_2 A_{1^{\prime }} A_{2^{\prime}}
 \ \left
[\tilde V_{1 1^{\prime }} \tilde V_{2 2^{\prime }}  J_{\Delta} +
V_{ 1 1^{\prime }, 2 2^{\prime }}  J_0 \right ]
\label{vertexz}
\eeqar
We abbreviate the residual dependence of the parton fields on the light
ray positions by indices $1,2$ for $A^{(+)}_1 (z_1)$, $ A^{(+)}_1 (z_2)
$ and by indices $1^{\prime }, 2^{\prime }$ for
$ A^{(-) \prime }_1 (z_{1^{\prime }}) $,
 $ A^{(-) \prime }_1 (z_{2^{\prime }}) $. The integration over the
transverse and $+$ components of $x_1, x_2$ is done due to the delta
functions and the integrals over the transverse and $+$ components of
$x_{1^{\prime}}, x_{2^{\prime }}$ are  represented by $ J_{\Delta},
 J_0$,
\beqar
c \ln \frac{Q^2}{m^2}  J_{\Delta} =  \int dx_{1^{\prime} +}
dx_{2^{\prime } + }
d^2 x_{1^{\prime } \perp} d^2 x_{2^{\prime } \perp } \partial^{\perp}_1
\partial^{ \perp * }_1  (x_{ 1 1^{\prime }}^2 )^{-1}
(x_{2 2^{\prime}}^2)^{-1} (x_{1^{\prime } 2^{\prime }}^2)^{-1}  \cr
c \ln \frac{Q^2}{m^2} \ 
 J_{0} =  \int dx_{1^{\prime} +}
dx_{2^{\prime }
+} d^2x_{1^{\prime } \perp} d^2 x_{2^{\prime } \perp } \partial^{\perp}_1
\partial^{ \perp * }_1  (x_{ 1 1^{\prime }}^2 )^{-1} (x_{2 2^{\prime
}}^2)^{-1} \delta^{(4)} (x_{1^{\prime } 2^{\prime }})
\eeqar
The integrals are regularized by taking into account the smearing of the
near light cone distribution by $\Delta $ and the large scale cutoff
$\delta$ for the other distribution. We shall do the integration in the
logarithmic approximation. For this we have substituted in
(\ref{vertexz}) already
the effective triple vertices omitting terms which do not result in
logarithmic integrals.

The integral over the transverse coordinates is logarithmic in the
region $Q^{-2} < |x_{\perp}^{\prime} |^2 < m^{-2} $.
\beqar
\label{J0calculation}
c \ln\frac{Q^2}{m^2} \ J_0  = 
\int {dx_+^{\prime} d^2 x_{\perp}^{\prime} \delta (z_{1\p 2\p}
\over
(-x_+^{\prime} z_{1 1\p} - |x_{\perp}^{\prime} |^2 + i\varepsilon)
(-x_+^{\prime} z_{ 2 2\p} - |x_{\perp}^{\prime} |^2 + i\varepsilon)
} = \cr
\int_{Q^{-2}}^{m^{-2}} {d^2 x_{\perp}^{\prime} \over
|x_{\perp}^{\prime}|^2 }
\int_{-\infty}^{\infty} d\bar \alpha
(\bar \alpha z_{11\p} + 1 - i \varepsilon)^{-1}
(\bar \alpha z_{22\p} + 1 - i \varepsilon)^{-1}
\delta (z_{1\p 2\p})
\eeqar
The result can be represented in the convenient form
$J_0 = J_0^{(0)}$, where
\beq
\label{J0p}
J_0^{(p)} (z_1,z_2;z_{1\p}, z_{2\p}) =
\int_0^1 d\alpha  \ \chi_0^{(p)} (\alpha) 
\delta (z_{11\p} - \alpha z_{12}) \
 \delta (z_{22\p} + (1 - \alpha) z_{12}),
\eeq
$$ \chi_0^{(0)} = 1, \chi_0^{(g)} = \alpha (1- \alpha). $$
The case of $J_{\Delta} $ is a bit more involved 
and can be done as follows.
\beqar
\label{Jdeltac}
c \ln {\frac{Q^2}{m^2}} J_{\Delta}  =
\int dx_{+1\p} x_{+2\p} d^2x_{\perp 1\p} d^2_{\perp 2\p}
(x^{\perp}_{1\p} \cdot x^{\perp}_{2\p} ) 
(-x_{+ 1\p} z_{11\p} - |x^{\perp}_{1\p}|^2 + i \ep )^{-2} \cr
(-x_{+ 2\p} z_{22\p} - |x^{\perp}_{2\p}|^2 + i \ep )^{-2}
(x_{+1\p 2\p} z_{1\p 2\p} - |x^{\perp}_{1\p 2\p}|^2 + i \ep )^{-1}
\eeqar
We use the Fock-Schwinger method to do the integrals over the $x_+$
variables and obtain
\beqar
\int_0^1 d\alpha_1 d\alpha_2 d\alpha_3 \delta(\sum \alpha_i -1)
\alpha_1 \alpha_2
\delta(\alpha_1 z_{1 1\p} - \alpha_3 z_{1\p 2\p} )
\delta(\alpha_2 z_{2 2\p}+ \alpha_3 z_{1\p 2\p} ) \cr
\int d^2x_{\perp 1\p} d^2_{\perp 2\p}
 (x^{\perp}_{1\p} \cdot x^{\perp}_{2\p} ) \ 
(\alpha_1 |x^{\perp}_{1\p}|^2 + \alpha_2 |x^{\perp}_{2\p}|^2  + \alpha_3
|x^{\perp}_{1\p 2\p}|^2 )^{-3}.
\eeqar
The logarithmic contribution of the transverse integral results in
\beqar
\int_0^1 d\alpha_1 d\alpha_2 d\alpha_3 \delta(\sum \alpha_i -1)
{ \alpha_1 \alpha_2 \alpha_3 \over (\alpha_2 + \alpha_3) (\alpha_1
\alpha_2 + \alpha_2 \alpha_3 + \alpha_3 \alpha_1 )^2 } \cr
\delta (\alpha_1 z_{11\p} - \alpha_3 z_{1\p2\p} ) \ 
\delta(\alpha_2 z_{2 2\p}+ \alpha_3 z_{1\p 2\p} ) \ \ 
\int_{Q^{-2}}^{m^{-2}} \frac{d|x_{1\p}^{\perp}|^2}{|x_{1\p}^{\perp}|^2}.
\eeqar
The convenient representation of the result is
$J_{\Delta} = J_{111} $, where
\beqar
\label{Jnnn}
J_{n_1 n_2 n_3} =
{\Gamma(n_1+n_2+n_3-1) \over \Gamma(n_1) \Gamma (n_2) \Gamma (n_3) }
\int_0^1 d\alpha_1 d\alpha_2 d\alpha_3 \delta(\sum \alpha_i -1) \cr
\alpha_1^{n_1 -1} \alpha_2^{n_2 -1} \alpha_3^{n_3 -1} \ 
\delta ( z_{11\p} - \alpha_1 z_{12} )
\delta( z_{2 2\p}+ \alpha_2 z_{1 2} ).
\eeqar
The general definition of the $\bar \alpha$ integrals appearing
here is
\beqar
J_{n_1 n_2 ...n_N} (x_1, x_2, ..., x_N) =
\int_{-\infty }^{+\infty} {d \bar \alpha \over 2 \pi i}
 [ \bar \alpha x_1 +1 - i \epsilon]^{-n_1}
[ \bar \alpha x_2  +1 - i \epsilon]^{-n_2} \cr
... [ \bar \alpha x_N +1 - i \epsilon]^{-n_N}.
\label{alphaint}
\eeqar
$n_i$ are positive integers. The Feynman parameter representation is
\beqar
J_{n_1 n_2 ...}(x_1, x_2, ...) =
{\Gamma (\sum n_i -1) \over \Gamma(n_1) \Gamma(n_2) ...\Gamma(n_N)}
\int {\cal D}^{(N)} \alpha \ \alpha_1^{n_1 -1}  \alpha_2^{n_2 -1}
\cr ...\alpha_N^{n_N -1} \
\delta(\sum \alpha_i x_i )
\label{alphafeynman}
\eeqar
In the latter integral the varibles $\alpha_i$
range from 0 to 1 and  ${\cal D}^{(N)} \alpha = d\alpha_1 ...d\alpha_N
\delta((1 - \sum \alpha_i ) $.

By decomposition into simple fractions one can reduce the number of factors
in the denominator of (\ref{alphaint}). In this way one derives relations
like
\beqar
J_{1 2}(x_1, x_2) = { x_1 \over x_1 - x_2} J_{1 1} (x_1, x_2), \cr
J_{1 1 1}(x_1, x_2, x_3) = { x_2 \over x_{23} } J_{1 1} (x_1, x_2)
 - { x_3 \over x_{23} } J_{1 1} (x_1, x_3) , \cr
J_{1 1 2}(x_1, x_2, x_3) = {- x_1 x_2 \over x_{13} x_{23} } J_{1 1} (x_1, x_2)
 + \left ({ x_2 \over x_{23} } + {x_1 \over x_{13}} \right ) J_{1 1 1} (x_1,
x_2, x_3 ) , \cr
J_{2 2 1}(x_1, x_2, x_3) = { x_3^2 \over x_{13} x_{23} } J_{1 1 1} (x_1, x_2,
x_3) \cr
 - {x_1 x_2 \over x_{12}^2 } \left ({ x_2 \over x_{23} } + {x_1 \over x_{13}}
 \right ) J_{1 1 } (x_1,x_2 ).
\label{relations}
\eeqar
We consider now in some detail the effective interaction of 
gluonic partons.  
The vertex in the external field effective action $ A^{* (+)} (x_1)
A^{* (+)} (x_2) A^{(-)} (x_{1^{\prime }}) A^{(-)} (x_{2^{\prime }}) $
referring to the case of parallel helicities (chiral odd  channel)
gets a contribution from the original quartic vertex  (\ref{action})
and
one from contracting two triple vertices by integrating over the quantum
fluctuation $A_q$.
The triple vertices in the Bjorken limit 
 corresponding to Fig. 2a, b are
\beq 
- i g {\partial^2_{1^{\prime}} \partial_1^{* \perp} \over
(\partial_1 \partial_{1^{\prime}}) \partial_1 } A_q^{a *} \
(A_1^* T^{a} A_{1^{\prime}}) \vert_{x_1=x_{1^{\prime}} = x_q}; \ \ \
 i g {\partial^2_{1} \partial_1^{ \perp} \over
(\partial_1 \partial_{1^{\prime}}) \partial_1 } A_q^{a} \
(A_{1}^* T^{a} A_{1^{\prime }}) \vert_{x_1=x_{1^{\prime}} = x_q}
\label{triplex}
\eeq
Writing down the extenal field vertex as in (\ref{vertexansatz}),
substituting the
asymptotic form of the external fields and doing the interations besides
of the ones in the light-ray coordinates we obtain as the particular
case of (\ref{vertexz})
\beqar 
\int dz_1 dz_2 dz_{1^{\prime}} dz_{2^{\prime}}
\left [ {\partial_1^2 \partial_{2^{\prime}}^2 + \partial_{1^{\prime }}^2
\partial_2^2 \over (\partial_1 + \partial_{1^{\prime}} )^2  \D_1 \D_2}
 J_{111} +
 {\partial_1 \partial_{2^{\prime}} + \partial_{1^{\prime }}
\partial_2 \over (\partial_1 + \partial_{1^{\prime}} )^2 }
 J_0  \right ]
\cr  (A_1^* T^{a} A_{1^{\prime}} )\
  (A_2^* T^{a} A_{2^{\prime}} )
\label{vertext1}
\eeqar
The first term in the brackets is  obtained by contracting the triple
vertices by substituting the points $1, 1^{\prime }$ in the second and
$2,2^{\prime } $ in the first vertex in (\ref{triplex}) and vice versa.
The
second terms emerges from the quartic vertex (\ref{action}) for the case
that
$A_1^*, A_2^*$ are of $A^{(+)} $ type and the remaining two field of the
$A^{(-)}$ type.

The second relation in (\ref{relations}) Fourier transformed to
light ray variables implies
\beq {\partial_1^2 \partial_{2^{\prime}}^2 \over (\partial_1 +
\partial_{1^{\prime}})^2 \D_1 \D_2 }  J_{111} \ + \
{ \partial_1   \partial_{2^{\prime}} \over (\partial_1 +
\partial_{1^{\prime}})^2 }  J_{0} \ = \  \partial_{1^{\prime }}
\partial_{2^{\prime}} \ \D_1^{-1} \D_2^{-1}  J_{1 1^{\prime }}^{(g)},
\eeq
where 
\beq
\label{J11p} 
 J_{1 1^{\prime }}^{(p)} =
 \int_0^1 d \alpha {\chi_1^{(p)} (\alpha) \over \alpha } \ 
\delta(z_{1 1^{\prime}} 
- \alpha z_{12}) \delta(z_{ 2 2^{\prime }}),
\label{J11g}
\eeq
$$ \chi_1^{(g)} = - (1-\alpha)^2, \chi_1^{(f)} = - (1-\alpha),
\chi_1^{(0)} = \alpha ( 1- \alpha ). $$
 
We use this relation and the one obtained by interchanging
$ 1, 1^{\prime } \leftrightarrow 2, 2^{\prime} $.
The remaining divergence signalled by ${1 \over \alpha} $ in the latter
integral  is cancelled after including the disconnected contribution of
order $g^2$ to the self energy in the two propagators connecting
$A_1^*$ with $A_{1^{\prime}}$ and
$A_2^*$ with $A_{2^{\prime}}$ separately,
\beq 
2 \int dz_1 dz_2 dz_{1^{\prime }} dz_{2^{\prime }} \
w_g \ \delta (z_{1 1^{\prime }}) \delta (z_{2 2^{\prime }})
\  (A_1^* T^{a} A_{1^{\prime}} )\
  (A_2^* T^{a} A_{2^{\prime}} ),
\label{disconnected}
\eeq
\beqar
w_p =  C_p \left ( \int_0^1 {d \alpha \over \alpha } \ + \ w_p^{(0)} \right ),
\ \ \ \ C_g = N, \ \ C_f ={N^2 -1 \over 2N}, \cr
w_g^{(0)} =  -\frac{1}{4} ({11 \over 3} - {2 N_f \over 3 N} ),
\ \ \ \ w_f^{(0)} = - {3 \over 4}.
\label{wp}
\eeqar
 Adding
(\ref{disconnected}) to (\ref{vertext1}) the
singular part of $w_g$ results in replacing in 
$ J^{(g)}_{1 1^{\prime }}$
(\ref{J11g}) ${1 \over \alpha } $ by ${1 \over [ \alpha]_+ } $, adopting the
conventional "plus prescription". 
In the following we understand the factor ${ 1 \over \alpha } $ in 
$J_{11\p}$ with this prescription.

We obtain
\beqar
 \int dz_1 dz_2 dz_{1^{\prime}} dz_{2^{\prime}}
[  J^{(g)}_{11^{\prime }} + w_g^{(0) }  \delta^{(2)} +
 (1, 1^{\prime } \leftrightarrow 2, 2^{\prime} ) ] \cr
 (\D_1^{-1}A_1^* T^{a} \partial A_{1^{\prime}} )\
  (\D_2^{-1} A_2^* T^{a} \partial A_{2^{\prime}} ), \ \ \ \ {\rm where } \
 \delta^{(2)}  \ = \  \delta (z_{1 1^{\prime
}}) \  \delta (z_{2 2^{\prime }})
\label{vertext}
\eeqar
We have calculated in the space-time effective action 
approach the vertex of parallel
helicity gluon interaction
to order $g^2$ in the logarithmic approximation. 

In the case of antiparallel helicity gluons the calculation of the
effective two-parton interaction results instead of (\ref{vertext1} )
in \beqar
\left ( {\partial_{1^{\prime }}^2 \partial_{2^{\prime }}^2 +
\partial_1^2 \partial_2^2 \over (\partial_1 + \partial_{1^{\prime}})^2 
\D_1 \D_2}
\  J_{111} +
 { (\partial_{1^{\prime }} \partial_{2^{\prime }} +
\partial_1 \partial_2)  \over
(\partial_1 + \partial_{1^{\prime}})^2 }
\  J_0 \right )
(A_1^* T^{a} A_{1^{\prime }} ) \ (A_2 T^{a} A_{2^{\prime }}^* )  \cr
 -  { ( \partial_{1} \partial_{1^{\prime }} +
\partial_2 \partial_{2^{\prime }})  \over
(\partial_1  + \partial_{2})^2 }
\ \  J_0 \ \
(A_1^* T^{a} A_{2} ) \ (A_{1^{\prime }} T^{a} A_{2^{\prime }}^* )  \cr
+ {(\partial_1 + \partial_{1^{\prime}} )^2 \over \D_1 \D_2} \  J_{111} \ \
(A_1^* T^{a} A_{1^{\prime }}^* ) \ (A_2 T^{a} A_{2^{\prime }} )
\label{vertexa1}
\eeqar
Here the self-energy contributions are not included yet. We separate in
the first bracket a term equal to the result for the parallel helicity
case (\ref{vertext1}). Adding now the self-energy contribution
(\ref{disconnected}) results in the replacement of the first bracket by
\beq
2 \partial_{1^{\prime }} \partial_{2^{\prime }}   [ J_{1 1^{\prime
}} + w_g^{(0)} \delta^{(2)} ] - (\partial_{1^{\prime }} -  \partial_1)
 (\partial_{2^{\prime }} -  \partial_2) \  J_{111} +
\partial_1 \partial_2    J_0 ,
\eeq
where now each term represents a regular operator.

The last two relations given in (\ref{relations}) for the $J$ integrals
Fourier transformed to light ray variables can be used now to substitute
$ (\partial_{1^{\prime }} -  \partial_1)
 (\partial_{2^{\prime }} -  \partial_2) \  J_{111} $ and
$ (\partial_{1^{\prime }} +  \partial_1)^2 \  J_{111} $
by $ J_{112}$ and $ J_{221}$ plus remainders involving
$ J_0$. This results in
\beqar
\label{antipg}
\left (2 [ J_{1 1^{\prime }}+  w_g^{(0)} \delta^{(2)} ]
+  J_{221} -2  J_{112}  \right  ) \
(\D_1^{-1} A_1^* T^{a} \partial A_{1^{\prime }} ) \ 
(\D_2^{-1} A_2 T^{a} \partial A_{2^{\prime }}^* )  \cr
+  J_{221} \ \
(\D_1^{-1} A_1^* T^{a} \partial A_{1^{\prime }}^* ) 
\ (\D_2^{-1} A_2 T^{a} \partial A_{2^{\prime }} )  \cr
 +
 {1  \over
(\partial_1  + \partial_{2})^2 }
 J_0 \ \  \{ -  (\partial_{1} \partial_{2^{\prime }} +
\partial_2 \partial_{1^{\prime }})
(A_1^* T^{a} A_{1^{\prime }}^* ) \ (A_2 T^{a} A_{2^{\prime }} ) \cr
 +  (\partial_{1} \partial_{1^{\prime }} +
\partial_2 \partial_{2^{\prime }})
[ (A_1^* T^{a} A_{1^{\prime }} ) \ (A_2 T^{a} A_{2^{\prime }}^* )
- (A_1^* T^{a} A_{2} ) \ (A_{1^{\prime }} T^{a} A_{2^{\prime }}^* ) ]
   \} .
\eeqar
Due to the commutation relation of the generators $T^{a} $ we have
for the gauge group brackets (\ref{brackets})
\beq
 (A_1^* T^{a} A_{1^{\prime }} ) \ (A_2 T^{a} A_{2^{\prime }}^* )
- (A_1^* T^{a} A_{2} ) \ (A_{1^{\prime }} T^{a} A_{2^{\prime }}^* )
=
(A_1^* T^{a} A_{2^{\prime }}^* ) \ (A_2 T^{a} A_{1^{\prime }} ).
\eeq
Thus the last  term in the brackets multiplying $ J_0 $  in 
(\ref{antipg}) is equal
to the first  one up to the sign and the exchange of $1^{\prime }$ and
$2^{\prime }$.
We remember that the above expressions are understood as integrated over
the light ray positions $ 1, 2, 1^{\prime}, 2^{\prime}$.  Therefore the
exchange of $1^{\prime }$ and $2^{\prime }$ is just a substitution of
integration variables. We conclude that the contribution involving
$ J_0$ cancels and arrive at the result (\ref{hgg}) up to
normalization.

The resulting effective action involves two types of fields $A^{\pm} (z) $
living on the light ray.
The (non-local) vertices in this action are converted into Hamiltonians by
considering $A^-$ as annihilation and $A^+$ as creation operators of partons.
The Bjorken asymptotics is calculated by studying the evolution 
generated by those Hamiltonians in the (Euklidian) time
variable $\xi $,
\beq
\xi (\kappa^2, m^2) = \int_{m^2}^{\kappa^2} {d \kappa^{\prime 2} \over
\kappa^{\prime 2} }\ { \alpha_S( \kappa^2) \over 2 \pi }.
\label{xi}
\eeq
We list the operators of the leading parton interaction in the QCD
Bjorken limit, the momentum kernels of which are well known,  in the
light-ray and Feynman parameter form. We restrict
ourselves to the ones involving gluons ($A$) and/or only one flavour and
one chirality of quarks ($f$). We use the abbreviations for the kernels
(\ref{J0p}, \ref{Jnnn}, \ref{J11p})  and also $J^{\pm} = z_{12}^{\pm 1} J $.
The position arguments of the fields will be abbreviated as indices 1,
$1^{\prime }$, 2, $2^{\prime }$. We suppress the label $(\pm)$ distinguishing
creation and annihilation operators, since the fields at points $1, 2$ act
always as creation and the fields at the points $1^{\prime }, 2^{\prime }$
always as annihilation operators.
The integration over the positions,
summation over colour indices and operator normal ordering is implied.

\noindent
parallel helicity interactions
\beqar
 \{ 4 [ J_{1 1^{\prime }}^{(g)} + w_g^{(0)} \delta^{(2)} ]
(\partial^{-1} A^*_1 T^{a} \partial A_{1^{\prime}})
-2 i [ J_{1 1^{\prime }}^{(f)} + w_f^{(0)} \delta^{(2)} ]
(\partial^{-1} f_1^* t^{a} f_{1^{\prime}} ) \} \cr
[-2 (\partial^{-1} A_2^* T^{a} \partial A_{2^{\prime}} ) +
i  (\partial^{-1} f_2^* t^{a} f_{2^{\prime}} ) ] \cr
-4i   J_{ 1 1^{\prime }}^{(0)}
 (\partial^{-1} f_1^{*} t^{\alpha} \partial A_{1^{\prime}})
\ ( \partial^{-1} A_2^* t^{* \alpha}  f_{2^{\prime }})   + {\rm c.c.}
\label{ttgf}
\eeqar
anti-parallel gluon interactions
\beqar
\{ 8 [  J_{1 1^{\prime }}^{(g)} + w_g^{(0)} \delta^{(2)} ]
\ + 4   J_{221} - 8  J_{112} \} \cr
 ( \partial^{-1} A_1^{*} T^{a} \partial A_{1^{\prime}})
( \partial^{-1} A_2 T^{a} \partial A^*_{2^{\prime }})   \cr
+ 4   J_{221}
 ( \partial^{-1} A_1^{*} T^{a} \partial A_{1^{\prime}}^*)
( \partial^{-1} A_2 T^{a} \partial A_{2^{\prime }})   + {\rm c.c.}
\label{hgg}
\eeqar
anti-parallel helicity quark interactions (one chirality, one flavour)
\beqar
\{-2  [ J_{1 1^{\prime }}^{(f)} + w_f^{(0)} \delta^{(2)} ]
+  J_{111} \}
 (\partial^{-1} f_1^* t^{a} f_{1^{\prime}} )
 ( \partial^{-1} f_2 t^{a} f_{2^{\prime }}^*)   \cr
+ 2  J_0^{(g)}
(\partial^{-1} f_1^* t^{a} f_{2})
 (\partial^{-1} f_{1^{ \prime}} t^{a} f_{2^{\prime }}^*) + {\rm c.c.}
\label{hff}
\eeqar
anti parallel helicity quark - gluon interactions
\beqar
-4i [J_{1 1^{\prime }}^{(g)} + w_g^{(0)} \delta^{(2)} ]
 ( \partial^{-1} A_1^{*} T^{a} \partial A_{1^{\prime}})
 \ ( \partial^{-1} f_2 t^{a} f_{2^{\prime }}^*)   \cr
-4i  [ J_{1 1^{\prime }}^{(f)} + w_f^{(0)} \delta^{(2)} ]
 ( \partial^{-1} f_1^* t^{a} f_{1^{\prime}})
( \partial^{-1} A_2 T^{a} \partial A^*_{2^{\prime }})  \cr
 - 4i   J_{211}
 [ ( \partial^{-1} f_1^* t^{  \alpha } \partial A_{1^{\prime}}^* ) \
  ( \partial^{-1} A_2 t^{ * \alpha} f_{2^{\prime }})
- (\partial^{-1} A_1^* T^{a}   \partial A_{1^{\prime }} )
(\partial^{-1} f_2 t^{a} f_{2^{\prime }}^* )  ] \cr
-8i   J_{111}
 (\partial^{-1} A_1^* T^{a}   \partial A_{1^{\prime }} )
(\partial^{-1} f_2 t^{a} f_{2^{\prime }}^* ) \ \
  + {\rm c.c.}
\label{hgf}
\eeqar
annihilation-type interactions
\beqar
-4i J_{111}^- \ (\partial^{-1} A_1^* t^{\alpha} f_{1^{\prime }}^* )
 (\partial^{-1} A_2 t^{*\alpha } f_{2^{\prime }} )
\ + \ 2i  \delta^{(2) -} \
 (\partial^{-1} A_1^* t^{* \alpha} f_{1^{\prime }} )
(\partial^{-1} A_2 t^{\alpha } f_{2^{\prime }}^* ) \cr
+ 12 i  J_0^{(g)-} \ (\partial^{-1} A_1^* T^{a} \partial^{-1} A_2)
 (f_{1^{\prime }} t^{a} f_{2^{\prime }}^* )  \cr
- \frac{2}{3} i  J_{221}^+ \
[ (\partial^{-1} f_1^* t^{\alpha} \partial A_{1^{\prime }} ) \
(\partial^{-1} f_2 t^{*\alpha } \partial A_{2^{\prime }}^* )
-
 (\partial^{-1} f_1^* t^{\alpha} \partial A_{1^{\prime }}^* ) \
(\partial^{-1} f_2 t^{*\alpha } \partial A_{2^{\prime }} ) ]  \cr
-i  J_{112}^+
[ (\partial^{-1} f_1^* t^{\alpha} \partial A_{1^{\prime }} ) \
(\partial^{-1} f_2 t^{*\alpha } \partial A_{2^{\prime }}^* )
\label{annihc}
\eeqar

\section{Conformal symmetry on the light ray}
\setcounter{equation}{0}

The conformal transformations act on the light ray positions $z$ as
$$ z \rightarrow z^{\prime } = { a z + b \over c z +d} $$
This action is generated by
\beq
 S^{(0),-} = \partial,  \ \ \ \  S^{(0) 0} = z \partial, \ \ \ \
S^{(0) +} = - z^2 \partial.
\eeq
The generators obey the
$sl(2)$ algebra,
\beq
  [S^0, S^{\pm} ] =  \pm S^{\pm}, \ \ \ [S^+, S^-] = 2 S^0
\eeq
We need other representations $\ell$ on functions of $z$ generated by
\beq
 S^{(\ell ) -} = S^{(0) -}, \ \ \
S^{(\ell) 0 } = z^{-\ell} S^{(0) 0} z^{\ell},
\ \ \ S^{(\ell) +} = z^{-2 \ell } S^{(0) +} z^{2\ell}.
\eeq
The representation space $V^{(\ell)}$ is spanned by $z^m$,
and the function
$1$ represents the lowest weight state.

The tensor product representation, $V^{(\ell_1)} \otimes V^{(\ell_2)} $,
is represented by polynomial functions of two variables
$ \psi(z_1, z_2)$ and is generated by
$ S_{12}^a  = S_1^{(\ell_1) a} + S^{(\ell_2) a} $.
It decomposes into
 irreducible representations, the lowest weight states of which,
$ S_{12} ^- \psi = 0,\ \  S_{12}^0 \psi = \lambda \psi $,
are $ \ \ \psi_n^{(0)} = (z_1 - z_2)^n,$ with $
  \lambda = \ell_1 + \ell_2 +n. $

The two-parton Hamiltonians obtained from QCD can be considered as acting on
the tesor product representation spaces
$$ \hat H : V^{(\ell_{1\p})} \otimes V^{(\ell_{2\p}) } \rightarrow
V^{(\ell_1)} \otimes V^{(\ell_2) } $$
The operator is symmetric if
\beq
 \left (S_1^{(\ell_1) a} + S_2^{(\ell_2)a} \right ) \hat H =
 \hat H \left (S_1^{(\ell_{1\p}) a} + S_2 ^{(\ell_{2\p})a} \right ).
\eeq
Writing the operator in integral form,
\beq
\hat H \psi (z_1,z_2) =  \int dz_{1\p} dz_{2\p} \ J(z_1,z_2|z_{1\p},z_{2\p} )
\psi(z_{1\p},z_{2\p}),
\eeq
this results in the following condition on the kernel,
\beq
\left ( S_1^{(\ell_1 a} + S_2^{(\ell_2) a} + S_{1\p}^{(1-\ell_{1^{\prime}})
a} +
S_{2\p}^{(1-\ell_{2^{\prime }}) a} \right )
J(z_1,z_2|z_{1\p},z_{2\p} ) = 0 .
\eeq
Here we assume that the integration is over closed contours in order to
have no boundary terms in the interation by parts.
The condition on the kernel is solved by the simple expression
\beq
J_{\ell_1,\ell_2,\ell_{1\p},\ell_{2\p}}(z_1,z_2|z_{1\p},z_{2\p} ) =
z_{12}^{a_{12} } \ z_{1\p2\p}^{a_{1\p 2\p}}
\ z_{11\p}^{a_{11\p} } \ z_{22\p}^{a_{22\p}}
 z_{12\p}^{a_{12\p} } \ z_{1\p2}^{a_{1\p2}},
\eeq
where the exponents are given in terms of the conformal weights $\ell_i$
and two paramenters $d, M$ as
\beqar
 a_{12} = d+1 - \frac 12 \sum \ell_i, \ \ \ a_{1\p2\p} = d -1 + \frac 12
\sum \ell_i \cr
 a_{12\p}  = - \frac 12 (\ell_1 + \ell_{1\p} -\ell_2 -\ell_{2\p}) +M, \ \ \
a_{1\p2} = + \frac 12 ( (\ell_1 + \ell_{1\p} -\ell_2 -\ell_{2\p}) +M \cr
 a_{11\p} = -( \ell_1 - \ell_{1\p} +1) -d-M, \ \ \
a_{22\p} = ( \ell_1 - \ell_{1\p} +1) -d-M
\cr
 \sigma = \ell_1 - \ell_{1\p} + \ell_2 - \ell_{2\p}
\eeqar
The general solution is a superposition of the above expression
with varying paramenters $M$ and $d$.
In order to close the contour one should choose a
double-loop enclosing two branch points as in Fig. 2  (Pochhammer contour).
For our applications we need integral operators where the integration is
along the real axis in $z_1\p$ and $z_2\p$.

\begin{figure}[htb]
\begin{center}
\epsfig{file=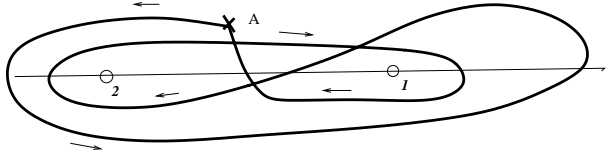,width=8cm}
\end{center}
\vspace*{0.5cm}
\caption{\small
Double-loop contour.
}
\end{figure}

The kernel for closed contours transforms into the kernel for integration
along the light ray by
 contraction of the contours to the branch cuts
and
rewriting the integral over finite intervals in $\alpha$
representation.

In the regular case, if the exponents are larger than -1, we obtain the
kernel in the form
\beqar
\label{regular}
z_{12}^{\sigma} \int_0^1 d\alpha_1 \int_0^{\alpha_1} d\alpha_2 \
\delta(z_{11\p} - \alpha_1 z_{12} ) \delta (z_{22\p} + \alpha_2 z_{12} ) \cr
\alpha_1^{a_{11\p}} \alpha_2^{a_{22\p}} (1-\alpha_1)^{a_{21\p}}
(1-\alpha_2)^{a_{12\p}} (1-\alpha_1 -\alpha_2)^{a_{1\p2\p}}
\eeqar
Singular cases
arise if some powers approach negative integers,
e.g. if $a = -1 + \varepsilon$ then the kernel is obtained from the one
above by by the
formal substitution
\beq
\label{singsubst}
 {1 \over \alpha^{1- \varepsilon} } \rightarrow \delta(\alpha) +
\varepsilon {1 \over [\alpha ]_+}.
\eeq
Now it is not difficult to check that the kernels obtained in QCD
have the conformally symmetric form with the weights
$\ell = \frac 32$ for gluons and $\ell = 1$ for quarks.

For example  if
$\ell_1 = \ell_2 = \ell_{1\p} = \ell_{2\p}= \ell $, $M = 0, d = \varepsilon $
we have in the integrand (\ref{regular})
\beq
 {(1-\alpha_1 - \alpha_2)^{\varepsilon + 2\ell -1} \over \alpha_1^{1 -
\varepsilon} \alpha_2^{1-\varepsilon} } \rightarrow
\delta  (\alpha_1) \delta(\alpha_2) + \varepsilon
\left ( \delta(\alpha_1) {(1-\alpha_2)^{2 \ell -1 } \over [\alpha_2]_+} +
\delta(\alpha_2) {(1-\alpha_1)^{2 \ell -1} \over [\alpha_1]_+} \right )
\eeq
The second term results in the  QCD  kernel
$$ J_{11\p}^{(2\ell -1)} (z_1,z_2|z_{1\p},z_{2\p} ) \ + \
( 11\p \leftrightarrow 22\p ), $$
$$
J_{11\p}^{(p) } = \int_0^1 d\alpha {(1- \alpha)^p \over \alpha }
[ \delta (z_{11\p} -\alpha z_{12} ) - \delta(z_{11\p}) ] \ \delta(z_{22\p}). $$
encountered in the results (\ref{ttgf} - \ref{annihc}).
One checks also easily that the functions $(z_1 - z_2)^n$ representing the
lowest weight states are eigenfunctions of this kernel. The eigenvalues are
\beq
\lambda_n = \int_0^1 d\alpha { (1-\alpha)^{2 \ell -1 +n } -1 \over \alpha} =
\psi(1) - \psi (2 \ell +n )
\eeq
They are proportional to the anomalous dimensions in the parallel helicity
exchange channel related to tranversity parton distributions.

The remaining kernels in the QCD operators correspond to the following
particular cases of the conformal ones:

$ gg \rightarrow gg $ (\ref{hgg}),

$ \ell_1 = \ell_2 = \ell_{1\p} = \ell_{2\p} =\frac 32 $

$M= 0, d = -1 $ results in $J_{112}$ and $M= 0, d= -2$ results in $J_{221}$.

$ f f \rightarrow ff $ (\ref{hgg}),

$ \ell_1 = \ell_2 = \ell_{1\p} = \ell_{2\p} =1$

$M= 0, d = -1$ results in $J_{111}$ and $ M= 0, d = -2+ \varepsilon$
results in $J_0^{(g)} $.

$ f f \rightarrow g g $ (\ref{annihc}),

$ \ell_1 = \ell_2 = \frac 32, \ell_{1\p} = \ell_{2\p} = 1,
 \sigma = -1 $

$M=0, d = - \frac 32  $ results in $z_{12} ^{-1} J_{111} $ and
$M = 0, d = - \frac 52 + \varepsilon $ results in $z_{12} ^{-1}
J_0^{(g)} $.

$ gg \rightarrow ff $, (\ref{annihc})

$\ell_1 = \ell_2 = 1, \ell_{1\p} = \ell_{2\p} = \frac 32,
\sigma = +1$

$M=0, d= -\frac 12 $ results in $z_{12} J_{112}$ and
$M=0, d= - \frac 32 $ results in $ z_{12} J_{221}$.

$ gf \rightarrow gf $ (\ref{hgf}),

$ \ell_1 = \ell_{1\p} = \frac 32 , \ell_2 = \ell_{2\p} =
1 $

$M = \frac 12, d= - \frac 32 $ results in $J_{111} - \frac 12 J_{211} $
and $ M = \frac 32, d= - \frac 32 + \varepsilon $ results in

$J_{1 1\p}  ^{(g)} + J_{2 2\p}^{(f)} $.

$ gf \rightarrow fg $ (\ref{hgf}),

$ \ell_1 = \ell_{2\p} = \frac 32, \ell_2 = \ell_{1\p} = 1
$

$M= 0 , d = - \frac 32 $ results in $ J_{121}$ and
$M= 0, d = - \frac 12+ \varepsilon $ results in $J_{1 1\p}^{(0)} $.

\section{Integrable systems of interacting parton }
\setcounter{equation}{0}

Consider operators acting on the tensor product
 $ V_1\otimes V_2 \otimes V_3 $ and in particular ones acting non-trivially
on two of the three spaces (as indicated by subscripts) and obeying
the Yang-Baxter equation (YBE)
\beq
R_{12} (u-v) R_{13} (u) R_{23} (v) \ = \
R_{23} (v) R_{13} (u) R_{12} (u-v).
\eeq
R- operators obeying this relation can be used to construct
integrable quantum systems. Consider $N$ subsystems, the quantum states of
which span the representation spaces $V^{(\ell_i)}, i=1, ..., N$ in a
periodic chain-like configuration where the interaction betweeen the subsystems
are due to exchanges along the chain with intermediate states out of the
auxiliary space $V^{(\ell_0)} $. A complete set of commuting operators
is obtained from the transfer matrix
\beq
\label{transfer}
t_{\ell_0} (u) = {\rm Sp}_{0} \left ( R^{(\ell_0, \ell_1)}_{01} (u)
R^{(\ell_0, \ell_2)}_{02} (u) ...  R^{(\ell_0, \ell_n)}_{0n} (u) \right )
\eeq
by expansion in the spectral paramenter $u$. In the homogeneous case,
$\ell_i = \ell, i=1, ..., N $, these operators appear as sums of local
operators (involving only a few neighbouring sites) for $\ell_0 = \ell$.
 The first non-trivial operator in the expasion
 can be considered as Hamiltonian
\beq
\label{hamiltonian}
 {d \over du} \ln t_{\ell} (u) |_{u=0} = \sum H_{i,i+1} ,
 \ \ \ \  \ H_{12} = P_{12} \ {d \over du} R^{(\ell_1,\ell_2)}(u)|_{u=0}
\eeq
Here $P_{12} $ denotes the operator of permutation of subspaces.

The best known example of a R-operator is the one related to the
Heisenberg XXX spin chain and is given by the $ 4 \times 4$ matrix
\beq
 R_{12}^{(-\frac 12, -\frac 12)} (u) = I_{4 \times 4} + \ (u + \frac 12)
\sigma^a \otimes \sigma^a
\eeq
In this case the representation spaces correspond to spin $\frac 12$
( $\ell = - \frac 12 $ in our notation) and are two dimensional.
The tensor product with the unit matrix acting on
the third subspace is not written explicitely here.

The
extension by $sl(2)$ symmetry leads to another solution, called also Lax
operator, appearing in
YBE for one subspace of arbitrary representation $\ell$ and two of the
fundamental one.
\beq
 R_{12}^{(-\frac 12, \ell )} (u) = I_{2 \times 2} \otimes \hat I^{(\ell)}
\ + \   (u + \frac 12)  \sigma^a \otimes \hat S_2^a
= (u +\frac 12 ) I + \hat R_2.
\eeq
The goal is to construct the universal R-operator: $ R^{(\ell_1,\ell_2)}$.
which appears in
YBE with $\ell_3= -\frac 12 $ and $ \ell_1, \ell_2 $ arbitrary.
This particular YBE  implies the defining condition conditions
\beqar
 (\hat R_1 + \hat R_2) R^{(\ell_1,\ell_2)}(u)  = R^{(\ell_1,\ell_2)} (u)
(\hat R_1 + \hat R_2), \cr
\left ( \frac u2 (\hat R_1 -\hat R_2) + \hat R_1 \hat R_2 \right )
R^{(\ell_1,\ell_2)} (u) =
R^{(\ell_1,\ell_2)} (u)
\left ( \frac u2 (\hat R_1 -\hat R_2) + \hat R_1 \hat R_2 \right )
\label{defconditions}
\eeqar
The first condition is equivalent to  the condition of conformal symmetry
which we solved before.
The second fixes the freedom left in (3.8) for conformal operators.

We look for the universal R operator in integral form. The conditions turn
into differential equations for the kernel
$R(z_1,z_2|z_{1\p},z_{2\p} )$.
\beqar
\left ( \hat R_1^{(\ell_1)} + \hat R_2^{(\ell_2)} +
\hat R_{1\p}^{(1-\ell_1)} + \hat R_{2\p}^{(1-\ell_2)} \right  )
R(z_1,z_2|z_{1\p},z_{2\p} ) = 0 \cr
\left (\frac u2 (\hat R_1^{(\ell_1)} -\hat R_2^{(\ell_2)}) +
 \hat R_1^{(\ell_1)} \hat R_2^{(\ell_2)} +
 \frac u2 (\hat R_{1\p}^{(1-\ell_1)} -\hat R_{2\p}^{(1-\ell_2)}) +
 \hat R_{2\p}^{(1-\ell_2)} \hat R_{1\p}^{(1-\ell_1)} \right )
R = 0
\label{kernelcond}
\eeqar
Here $\hat R^{(\ell)}_i $ is the non-trivial piece in the Lax operator
matrix acing on $z_i$.
$$
 \hat R_i =
\left (
\begin{array}{cc}
S_i^{(\ell)0 } & S_i^{(\ell)-} \\
S_i^{(\ell) +} & -S_i^{(\ell) 0} \\
\end{array}
\right )
= \left (
\begin{array}{c}
1 \cr
z_i \cr
\end{array}
\right )
\ (z_i \ \ -  1) \partial + \ell \left [ \left (
\begin{array}{c}
0 \cr
1 \cr
\end{array}
\right ) \ ( z_i  \ \ \ -1 ) +
\left (
\begin{array}{c}
1 \cr
z_i \cr
\end{array}
\right )
\ (1 \ \ 0 ) \right ].
$$
The decomposition of the $2 \times 2$ matrix structure into
simple tensor products simplifies essentially the solution of
(\ref{kernelcond}).
As an equivalent representation we may take the matrix element
with repect of two vectors paramentrized by positions $z_l$ and $z_r $.
\beqar
< l | R_i | r >  =
(z_l \ \ \  -1)  \hat R_i \ \
\left (
\begin{array}{c}
1 \cr
z_r \cr
\end{array}
\right )
=\cr
-z_{li} z_{ir} \partial_i - \ell_i ( z_{ir} - z_{li} ) =
z_{ir}^{-\ell_i +1} z_{li}^{\ell_i +1} \ \partial_i \
z_{ir}^{\ell_i } z_{li}^{-\ell_i }
\eeqar
We rewite the conditions (\ref{kernelcond})
in terms of those generic matrix elements and
use the fact that some derivative terms vanish for if $z_l$ or $z_r$
tends to some of the points $z_i$. In this way one obtains that
 the conditions (\ref{kernelcond}) are equivalent to
four first order equations.
\beq
 {\cal D}_i \ R (z_1,z_2|z_{1\p}, z_{2\p} ) = 0.
\eeq
Their consistency implies that the first order differential operators are
similarity transformations of the simple derivatives,
$$ {\cal D}_i = R \partial_i R^{-1}, $$
and this transformation is just given by the wanted
R operator kernel,
\beq
R (z_1,z_2|z_{1\p}, z_{2\p} ) =  z_{12}^{a_{12}} \ \ z_{1\p2\p}^{a_{1\p2\p}}
\ \ z_{12\p}^{a_{12\p}} \  \ z_{12\p}^{a_{12\p}}.
\eeq
It is indeed a
particular case of the conformal symmetric kernels with
$\ell_1 = \ell_{1\p}, \ell_2 = \ell_{2\p}, d=u, M = -u-1 $,
\beqar
 a_{12} = u+1 - \ell_1 - \ell_2, \ \ \ a_{1\p2\p} = u -1 + \ell_1 + \ell_2
\cr
 a_{12\p} = -u- 1 - \ell_1 + \ell_2, \ \ \ a_{1\p2} = -u -1 + \ell_1 - \ell_2
\cr
 a_{11\p} = a_{22\p} = \sigma = 0.
\eeqar
In the special case
$\ell_1 = \ell_2 $
we have
$$ R^{(\ell,\ell)} = \int _0^1 d\alpha_1 \int_0^{\alpha_1} d\alpha_2
\delta (z_{11\p} - \alpha_1 z_{12} ) \delta (z_{22\p} + \alpha_2 z_{12} ) $$
$$
(1-\alpha_1 -\alpha_2)^{u-1+2 \ell} (1-\alpha_1)^{-u-1} (1-\alpha_2 )^{-u-1} $$
We expand for small $u$ using (\ref{singsubst})
$$ \delta (z_{21\p}) \delta (z_{12\p}) + $$
 $$
u \int_0^1 d\alpha {(1-\alpha )^{2\ell -1} \over [\alpha]_+ }
\delta (z_{12\p} - \alpha z_{12} ) \delta (z_{21\p}) \ + \
( 11\p \leftrightarrow 22\p ) + {\cal O} (u^2) $$
Comparing with the formula for the integrable Hamiltonian (\ref{hamiltonian} )
we see that
the terms with $J_{1 1\p}^{(g/f)} $ in the QCD operators (\ref{ttgf} -
\ref{annihc})
coincide with this Hamiltonian for $\ell_g = \frac 32 $ and $\ell_f = 1$.

It is remarkable that particular cases of parton interaction operators
coincide with the integrable Hamiltonian of a generalization to
(non-compact) representations $\ell = 1$ or $\frac 32$ of the
periodic homgeneous XXX spin chain. The multi-parton exchange contribution
for parallel helicities of only gluons or only quarks is determined just by
this operator. In the approximation of a large number of colours $N_C$
 closed chain configurations of gluons dominate and the corresponding
contribution to the Bjorken asymptotics is calculated from the solution of
the XXX integrable system. In particular the lowest energy eigenvalues
determine the powers of the large scale $s$ of this contributions.

Integrable systems corresponding to other helicity configurations
of QCD parton interactions have not been constructed yet. There are some
results related to open chains with three sites \cite{BDKM}.
Besides of the
terms related to the integrable Hamiltonian $J_{11\p}^{(q/g)} + (1
\leftrightarrow 2) $
there are other terms in the QCD results.
However there is a close relation of the other operator terms appearing
in QCD to the R operator.

For the gluon case (\ref{hgg}) $ P_{12} R^{(\frac 32, \frac 32)} (u) $
for $u= -1 $ results in the kernel $J_{112}$  and for $u=-2$ in $J_{221}$.
For the fermion case (\ref{hff}) $ P_{12} R^{(1,1)} (u) $ results for
$u= -1$ and $-2$ is $J_{111}$ and $J_0^{(g)}$.
The kernels of the annihilation interactions (\ref{annihc}) are obtained as
$z_{12}^{-1} P_{12} R^{(1,1)} (u) $ with $u = -1, -2$ for $f f \rightarrow
gg $
and $z_{12} P_{12} R^{(\frac 32, \frac 32)} (u) $ with $u = -1, -2$
for $ g g \rightarrow f f $.

In case of mixed quark-gluon interactions $P_{12} R^{(1, \frac 32)} (u) $
with $u= -\frac 12 $ reproduces $J_{211}$ and by crossing also
$J_{111}- \frac 12 J_{121} $ and with $u= -\frac 32 $ reproduces
$J_{1 1\p}^{(0)} = J_{1 1\p}^{(f)} -J_{1 1\p}^{(g)} $.
In this case there is one contribution left, the kernel
$ J_{1 1\p}^{(f)} + J_{1 1\p}^{(g)} $, not reproducible from this
R-operator kernel.
 In the supersymmetric extension this kernel
appears as a component of the interaction Hamiltonian of XXX chain
constructed from the superconformal $sl(2|1)$ solution of the
Yang-Baxter equation \cite{DKK}.

{\large \sl Acknowledgements }

I am grateful to S. Derkachov and D. Karakhanyan for collaboration
on topics related to this contribution.
 I thank the organizers of the
36th Winter school of St. Petersburg Nuclear Physics Institute
for kind hospitality.
The work has been supported in part by the German Federal Ministry
BMBF.

\end{document}